\documentclass[english,doublespacing]{elsart}
\usepackage[T1]{fontenc}
\usepackage[latin1]{inputenc}
\usepackage{babel}
\usepackage{graphics}
\usepackage{subfigure}

\makeatletter

\providecommand{\LyX}{L\kern-.1667em\lower.25em\hbox{Y}\kern-.125emX\@}

\usepackage{amsfonts}

\makeatother

\journal{Physica C}
\begin{document}
\begin{frontmatter}

\title{Electrical characterization of YBCO single crystal surfaces oriented
in any crystallographic direction}

\author[crtbt]{H. Guillou\corauthref{hg}}
\corauth[hg]{Corresponding author : phone : 33 4 76 54 9591 \\ fax : 33 4 76 54 9425}
\ead{herve.guillou@ujf-grenoble.fr}
\author[crtbt]{J. Chaussy}
\author[crtbt]{M. Charalambous}
\author[NCRS]{M. Pissas}

\address[crtbt]{CNRS CRTBT, 25 av. des Martyrs, BP166X, 38042 Grenoble Cédex, France.}
\address[NCRS]{Institute of Materials Science, NCRS Demokritos 15310 Ag. Paraskevi,
Athens, Greece.}

\begin{abstract}
Although considerable studies have been carried out, the true nature
of high-T\( _{c} \) superconductors (HTCS) is still not clear. Pseudogap
phase at high temperature as well as possible time reversal symmetry
breaking at low temperature need further investigations. The need
of carefully made samples showing the intrinsic properties of superconductivity
is essential to test new theoretical developments. We present in this
paper how to control crystallographic orientation in the junction
and a technique developed to determined the quality of the interface
barrier between a gold electrode and a HTCS~: YBa\( _{2} \)Cu\( _{3} \)O\( _{7-\delta } \).
This potentially allows us to perform Andreev spectroscopy in the CuO\( _{2} \)
planes of cuprate superconductors as a function of temperature, crystallographic
orientation and doping.
\end{abstract}
\end{frontmatter}

\section{Introduction}

Since their discovery by Bednorz and M\"{u}ller \cite{bednorz86},
high-T\( _{c} \) superconductors have been the subject of considerable
studies. However, their complex structure and properties are still
puzzling solid state physicists~\cite{orenstein00}. One unanswered question 
of importance is the relation between the superconducting gap
and the pseudogap often seen in ARPES~\cite{loeser96,pavuna00} and
tunneling spectra~\cite{renner98}. Their evolution both in amplitude
and symmetry with doping and temperature is of great interest and
needs to be determined. Low temperature tunneling studies also showed a
possible break of time reversal symmetry~\cite{sharoni01,convington97,fogelstrom97}
as well as zero bias anomalies~\cite{Hu1994,alff98,lesueur91,aprili99,krupke99} whose
origin is not clearly established~\cite{walker99,samokhin01}. Provided
one can distinguish a pseudogap from a superconducting gap, angularly
resolved tunneling experiments are good candidates to study these
issues. In fact planar junctions are mechanically and chemically stable
over the broad range of temperatures necessary to investigate these
properties. Moreover, recent arguments~\cite{deutscher99} suggest
that controlling the interface's potential and thus the transport
mechanism could allow to determine independantly the energy scales
relevant for pairing and phase stiffness~\cite{emery95} in materials
where the superconducting density is low such as in HTSC. Thus, the
realization of angularly resolved planar contacts with controlled
transparency on high-\( T_{c} \) materials could propose experimental
answers to still unanswered questions on HTCS. In this article we present
a technique that allows us to realize planar junctions with controlled
orientation and transparency on YBCO single crystals. We show how to
control and characterize the interface quality. Briefly, the junctions
are realized on the side of the crystals and allow direct injection
of uncoherent quasi-particles in the copper-oxide planes. The orientation
of the junction's plane is set by mechanical polishing of the crystal
and the surface barrier is controlled using ion polishing. The junction
is geometrically defined using photolithography and conductance spectra
\( dI/dV(V) \) are measured down to \( T=5 \)K using a lock-in technique. 
The measurements, carried out on several crystals prepared with
different orientations and interface potential, are presented in the
third section and interpreted using a phenomenological model. This
experiment has an essential advantage~: metallic interfaces,  obtained
after ion polishing,  allow us to perform Andreev spectroscopy
on YBCO/Au crystallographically oriented planar junctions.

\section{Experimental}

\begin{figure}
{\centering \subfigure[]{\resizebox*{0.4\textwidth}{!}{\includegraphics{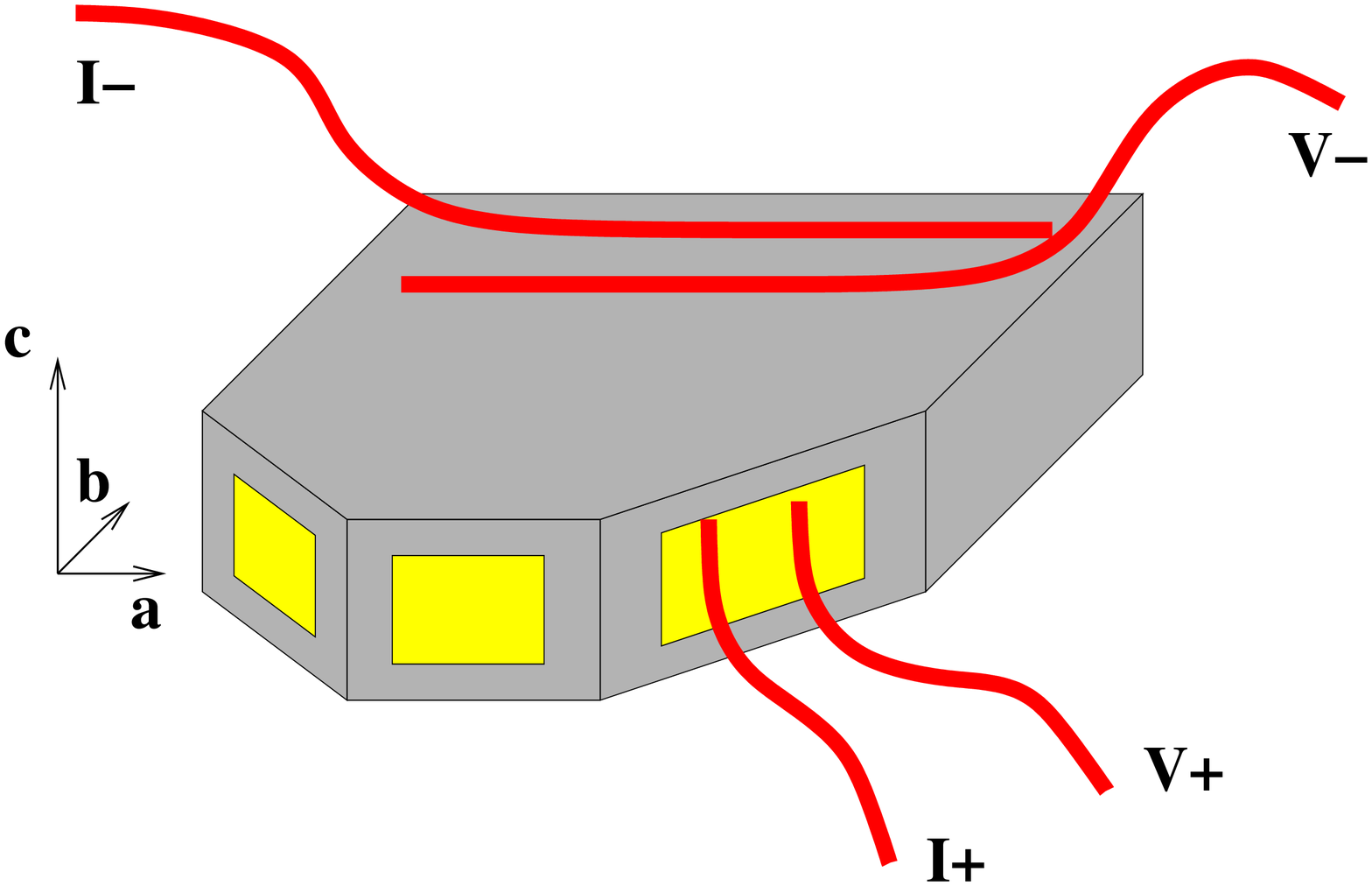}}}\subfigure[]{\resizebox*{0.4\textwidth}{!}{\includegraphics{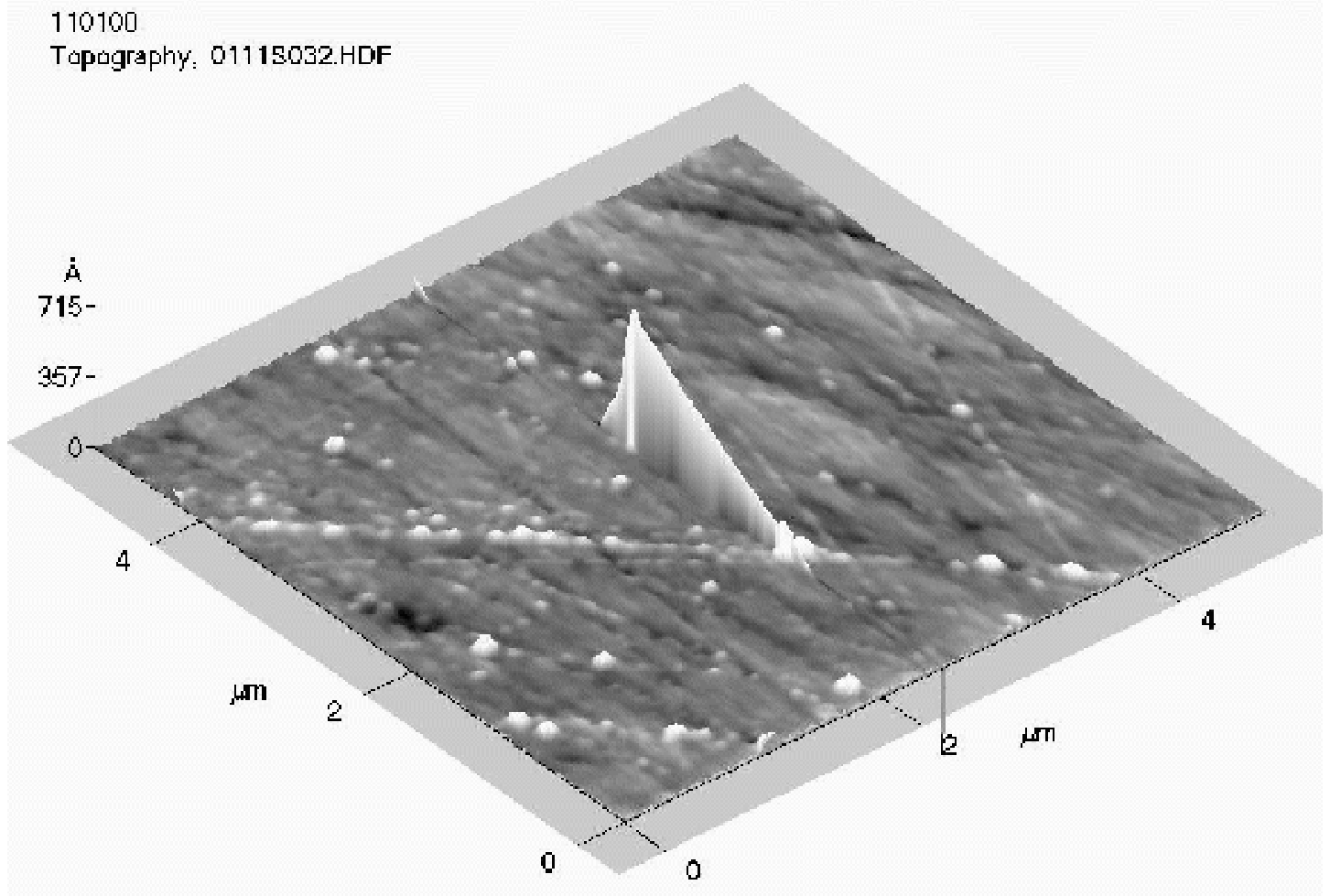}}}\par}

\caption{\label{fig1}The junctions are realized on a plane containing
the \protect\( c\protect \)-axis of the crystal (left). The orientation
of the surface is controlled and determined by mechanical polishing
that yields very small roughness as shown with AFM (right). (a) Planar
junctions on the \char`\"{}side\char`\"{} of YBCO single cristal.
(b) AFM scan of the mechanically polished surface. The RMS roughness
is about 1--2 nm.}
\end{figure}

The YBa\( _{2} \)C\( _{3} \)O\( _{7-\delta } \) (YBCO) single crystals
used in this work were optimally doped with a \( T_{c} \) of 93~K
and a transition width of 0.3~K as measured by ac-Hall susceptibility.
The typical crystal size is \( 500\mu \rm {m}\times 300\mu \rm {m}\times 80\mu \rm {m} \).
Crystals with the flattest surfaces and the sharpest corners were
selected to be polished. The backleads \( I_{-} \) and \( V_{-} \)
are attached to the crystal's rear surface by diffusing silver epoxy
at \( 550^{\circ } \)C for one hour in an atmosphere of flowing O\( _{2} \) 
as shown in figure~\ref{fig1}a). This ensure mechanically
robust and electrically excellent contacts.  

Since the junctions will be realized so as to inject current directly
in the CuO\( _{2} \) planes the junction plan must include the \( \overrightarrow{c} \)
axis. In orther words, we must be able to cut the crystal in a controllable
manner in order to expose a surface less than 100 \( \mu  \)m thick
on its side. Polishing seems the simplest solution. The sample is embedded
in an epoxy matrix in order to minimize  side effects during polishing.
Prior to casting, the crystal is oriented and attached to a copper
disk to ensure a good thermal contact. The orientation is controlled
within a couple of degrees with respect to the CuO\( _{2} \) planes
and within \( \pm 7^{\circ } \) with respect to the c-axis. This
uncertainty is much lower than the current injection cone, usually estimated
to be around \( \pm 15^{\circ } \)~\cite{wolfbook} for low transmission
interfaces. After polymerization, the whole epoxy block is carefully
polished with fine diamond paste. Great care has been taken in order
to avoid chemical attack of crystals (water-free polishing), cracks
(low speed spinning), diamond inclusions and rounded surfaces (low
pressure). Between each stage the sample is washed in an ultrasonic
bath and rinsed with pure acetone and ethanol to avoid contamination.
The resulting surface is flat and without any scratches as observed with
a DIC microscope. Some surfaces were observed using AFM and the measured surface
roughness was  found to be around 15~\AA and
sample independant. A typical image is shown on the figure \ref{fig1}b).
The scanning size did not  significantly change the measured roughness.
It has been pointed out by Walker \emph{et al.}~\cite{walker99}
that diffuse scattering at the interface can significantly complicate
the transmission process. Thus it is crucial to keep the roughness
comparable with the Fermi wavelength of the charge carriers in YBCO.
We note that very little care has yet been taken  to communicate the
structural properties of the surface on which planar junctions are
elaborated, eventhough a parameter such as roughness is crucial to
predict the transport properties of junctions. The measured roughness
is only twice the carrier wavelength and although some charge
carriers probably undergo diffuse scattering, polishing of YBCO surface
yields outstanding results concerning the surface's morphology. 

At this stage the flat and smooth crystal surface is ready to be used
to elaborate the junctions. However, polishing induces a fine amorphous
or deoxygenated layer directly below the surface. Since we are concerned
with conserving the bulk properties of superconductivity throughout the 
sample, this layer must be removed. Ion beams have been used by several
groups to smooth YBCO films~\cite{hebard89}, to improve superconducting
properties of epitaxial YBCO films~\cite{chen97} and to pattern,
without damage, YBCO thin films~\cite{xavier94}. In line with these
results we used low energy (300 \( e \)V) Xe ion beams at a normal
incidence to polish and improve the superconducting properties of
the crystal surface. The polishing efficiency is related to the sputtering
yield of the impinging ions. This has recently been reviewed~\cite{carter01}
and it appears that for normally incident ions, the sputtering yields
are maximum for a surface oriented roughly \( 60^{\circ } \) off the
normal. Therefore this treatment  erodes  these asperities
and improves the surface roughness, as well as removes the degraded
surface layer. It is well known that the superconducting properties
of cuprates are very sensitive to oxygen contents. During the ion
beam polishing a large amount of energy is dissipated in the surface's
vicinity. To avoid heating and further loss of oxygen during the ion
beam polishing the crystal is maintained at liquid nitrogen temperature.
After ion polishing, a \( 150 \) nm thick gold layer is sputtered
\emph{in-situ} and will be used as a counter electrode. 

\begin{figure}
{\centering \resizebox*{0.4\textwidth}{!}{\includegraphics{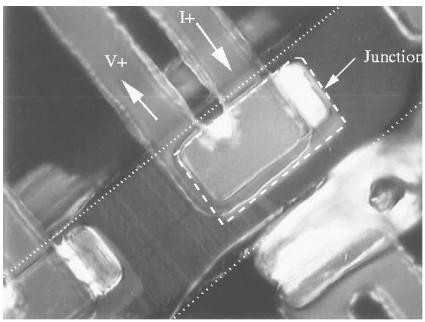}}}

\caption{\label{fig2}Contact and probes geometry. The crystal boundaries
are underline by the dotted lines. The junction is realized within
the discontinous line. The contact nominal surface is \protect\( S=30\mu \rm {m}\times 60\mu \rm {m}\protect \).}
\end{figure}

We are interested in the transport properties of the YBCO/Au interface.
The geometry of the junction and the leads that are attached to it
are defined using a photolithographic process and are shown in figure~\ref{fig2}.
The nominal junction surface is \( S=0.18\cdot 10^{-3} \)~cm\( ^{2} \).
The gold resistivity is of the order of \( 0.1\, \, \mu \Omega \cdot \rm {cm} \)
at low temperature.  For a uniform current injection, this requires  an interface
resistance higher than \( 4\cdot 10^{-6}\, \, \Omega \cdot \rm {cm}^{2} \)
for the nominal junction dimensions~\cite{rzchowski2000,vandeveer97,pedersen67}.
This condition was always fulfilled and the only source of inhomogeneities
in the current injection is the inhomogeneity of the superconductor's
surface. 

The measurement temperature can be varied from 5~K to 150~K, the
current is measured as a function of the bias voltage and the slope
of the \( I(V) \) characteristic is measured using a lock-in technique.
The experimental setup gives an absolute value of the dynamic conductance
as shown in the inset in figure~\ref{fig5}a).

In this article we stress the fact that the transparency of the YBCO/Au
interface can be controlled by choosing the method of preparation of the 
junctions.
In fact we compare in the next section the results obtained on crystals
prepared differently. Two behaviours were observed depending on the
preparation technique~: i) the gold counter-electrode is directly
sputtered after the \emph{mechanical} polishing. This leads to an
insulating behaviour of the interface resistance and thus a tunnel
effect at low temperature. ii) The counter-electrode is evaporated
\emph{in-situ} after low temperature and  low energy \emph{ion} polishing.
This leads to a metallic behaviour of the interface resistance and
hence Andreev effect at low temperature.

\section{Results and discussion}

When considering an interface between two materials, the transmission
coefficient can be estimated in principle from the absolute value
of the surface resistivity. For  high temperature superconductors in
contact with a normal metal this is estimated~\cite{deutscher91,kupriyanov91}
to be \( R_{c\Box }=(\hbar /e^{2})\times 2\pi ^{2}/Dk^{2}_{F} \)
where \( R_{c\Box } \) is the measured surface resistivity, \( k_{F} \)
the Fermi wave vector of carriers in the superconductor and \( D \)
the transmission coefficient. For an ideal interface where \( D=1 \)
and a Fermi wave vector \( k_{F}\sim 10^{-8}\, \rm {cm}^{-1} \) the
estimation gives \( R_{c\Box }\sim 10^{-11}\, \Omega \cdot \rm {cm}^{2} \).
Numerous experiments have been designed to reach this ideal limit
but the best results \cite{tsai90,jing89} gave interface resistances
100 times larger than the theoretical estimate. To explain this discrepency
Mannhart \emph{et al.}~\cite{mannhart99} proposed the existence
of a depletion layer near the interface that tends to enhance the
interface resistance. Moreover spatially resolved measurements of
the proximity effect~\cite{koyanagi92} induced in  thin gold layers
showed large inhomogeneities. More recent STS measurements~\cite{cren00a}
on high quality Bi\( _{2} \)Sr\( _{2} \)CaCu\( _{2} \)O\( _{8-\delta } \)
thin films showed quasi-particles spectra going from S-I-N tunneling
to Semi-C-I-N on a scale of a few nm. These results show that calculating
the transmission coefficient from resistance measurements of macroscopic
interfaces is difficult  mainly because of intrinsic large inhomogeneities
on high temperature superconductor surfaces. 

\begin{figure}
{\centering \resizebox*{0.4\textwidth}{!}{\includegraphics{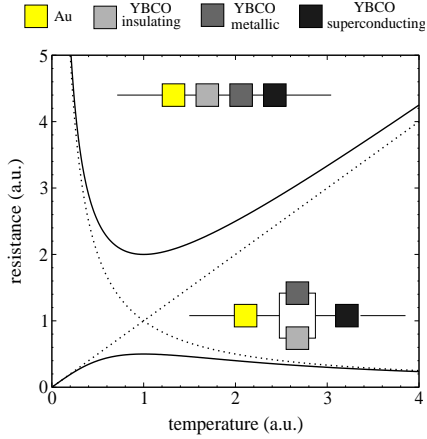}}\par}

\caption{\label{fig3}Modeling of the interface between YBCO and a normal
metal. Top: all the layers are in series. The resistance vs T has
positive curvature: \protect\( R(t)=\alpha (1/t)+\beta t\protect \).
Bottom: the insulating and metallic layers are in parallel. The resistance
vs T has a negative curvature: \protect\( R(t)=1/(\alpha (1/t)+\beta t)\protect \).}
\end{figure}

\begin{figure}
{\centering \subfigure[]{\resizebox*{0.4\textwidth}{!}{\includegraphics{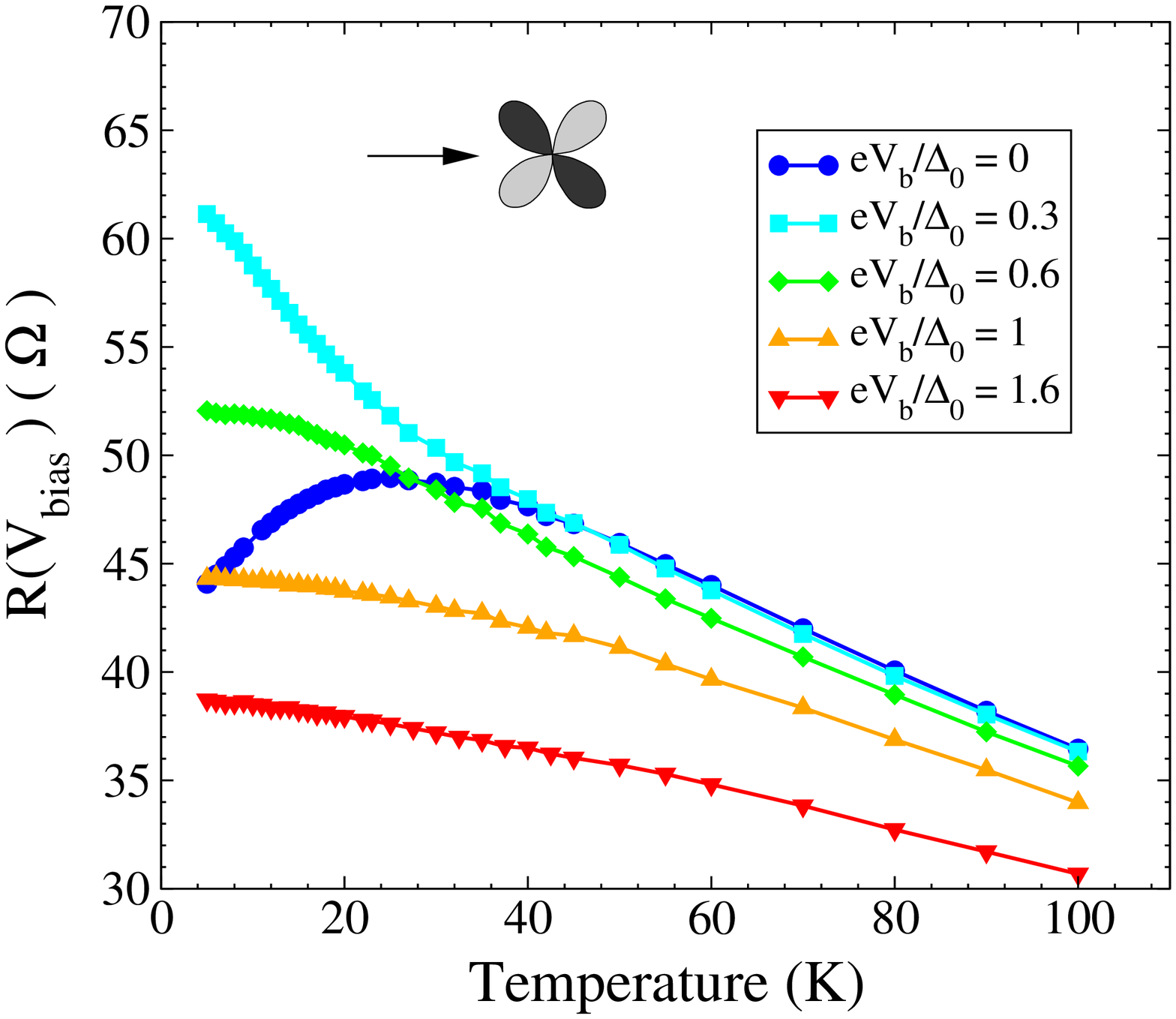}}}\subfigure[]{\resizebox*{0.4\textwidth}{!}{\includegraphics{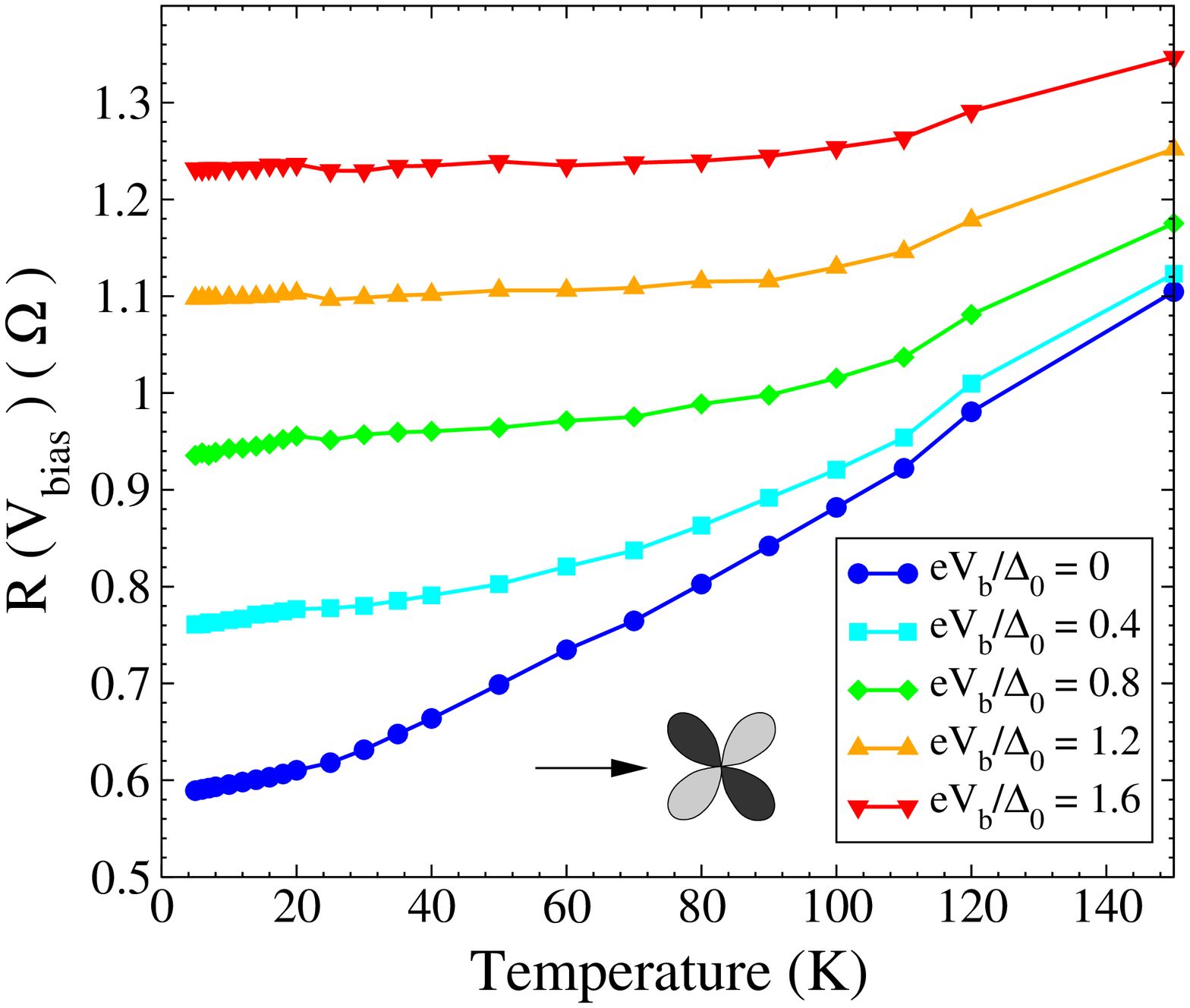}}}\par}
\caption{\label{fig4}Evolution with temperature of the resistance for a
bias voltage \protect\( V_{b}\protect \). The junctions are realized
on a (110) oriented surface. The two different behaviours observed
are related to the preparation techniques. (a) Insulating (tunnel)
limit obtained when the crystal is only mechanically polished. (b)
Metallic regime obtained after ion polishing of the crystal's surface.}
\end{figure}

In order to determine the dominant transport mechanism through the interface,
the sole knowledge of the interface resistance is insufficient. However
its temperature dependance provides indications on which transport
mechanism dominates. Indeed, it is reasonable to describe the junctions
as an insulating and a metallic layer. If the two layers are in serial
an insulating behaviour is expected for low enough temperatures. On
the contrary if  the two layers are in parallel a metallic behaviour is
expected for low enough temperatures as shown in fig.~\ref{fig3}).
From these properties we deduce the dominant transport mechanism
through the interface. At low temperatures, a metallic interface will
show Andreev reflections whereas an insulating interface will show tunneling
effects.  Fig.~\ref{fig4}) shows the temperature dependance
of the interface resistance for different bias voltages for the two
preparation procedures for a (110) oriented surfaces. 

For  fig~\ref{fig4}a) the surface of the crystal was simply
mechanically polished and carefully cleaned. The junction resistance
increase for decreasing temperatures is characteristic of insulating
properties. Tunneling will be the dominant transport mechanism through
the interface at low temperatures. We note that the behaviour of the
resistance at zero bias is different from the ones at finite bias.
The increase of conductivity for low bias is the signature of a Zero
Bias Conductance Peak (ZBCP). ZBCP's in HTSC are explained
either in the Andreev Bound States~\cite{Hu1994} mechanism, or in
the presence of surface defects that create Impurity Bound States~\cite{samokhin01}.
Although these different origins can in principle be differentiated
by their different magnetic field properties, their probable coexistence
does not facilitate the determination of their contribution to the
global ZBCP. The curves in fig.~\ref{fig4}b) show a clear metallic
behaviour  of the junction's resistance with temperature. The junction
was made in situ after ion polishing. The metallic behaviour comes
from the removal of the degraded layer at the surface by the ions.
 Thus at low temperature we expect to measure mostly Andreev transport
through the interface. 

\begin{figure}
{\centering \subfigure[]{\resizebox*{0.4\textwidth}{!}{\includegraphics{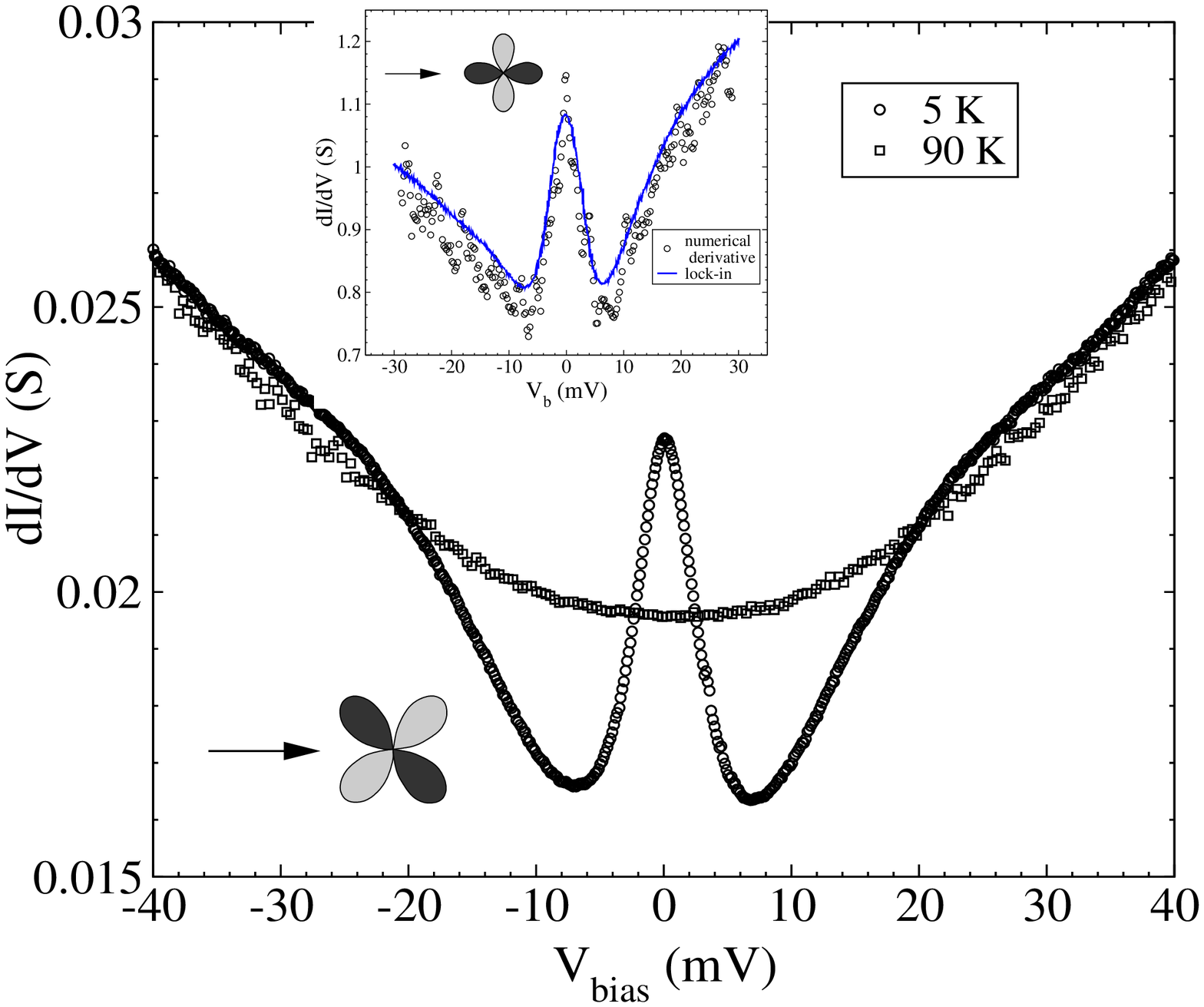}}}\subfigure[]{\resizebox*{0.4\textwidth}{!}{\includegraphics{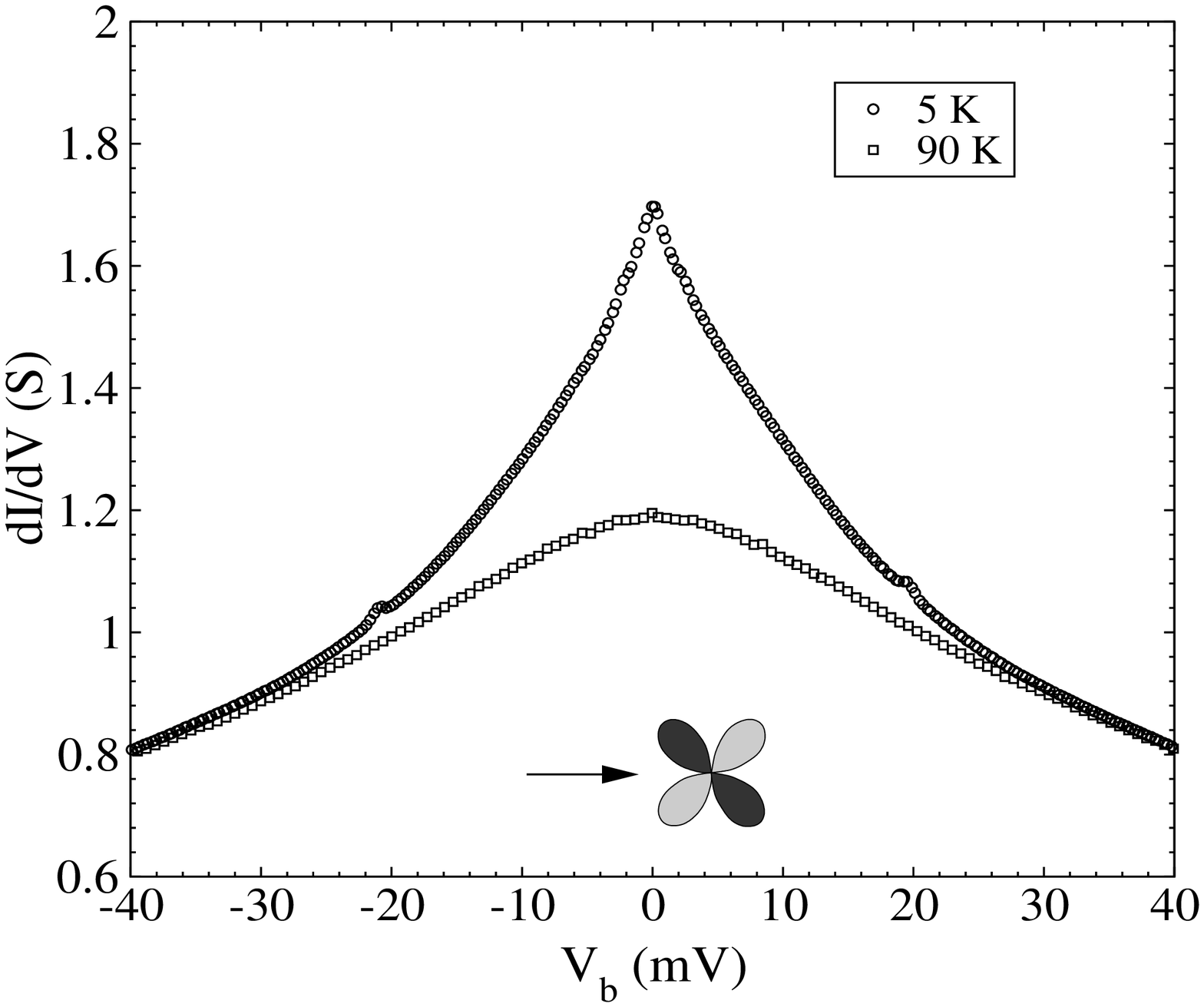}}}\par}
\caption{\label{fig5}Dynamic conductance spectra obtained for YBa\protect\( _{2}\protect \)Cu\protect\( _{3}\protect \)O\protect\( _{7-\delta }\protect \)(110)/Au
junctions for differents transport regime through the interface. The
spectra measured at 90 K were shifted by a constant to overlay the
spectra measured at 5 K. The absolute values can be determined from
the fig.~\ref{fig4}). (a) tunneling spectrum obtained at low temperature;
inset : comparaison of the measured dynamic conductance and the one
computed from the numerical derivative of the \protect\( I(V)\protect \)
curve (not shown). (b) Andreev spectra obtained after ionic polishing
of the cristal surface}
\end{figure}

The fig.~\ref{fig5}) shows the measured spectra for the tunneling
and the Andreev limit obtained on different crystals. For the two
junctions shown here the current injection was along the line of node
of a \( d_{x^{2}-y^{2}} \) order parameter. In the tunneling limit
(fig.~\ref{fig5}a), the spectra show all the expected qualitative
features of a \( d \)-wave superconductor : \emph{i)} the DOS is
depreciated below an energy of about 25 m\( e \)V corresponding to
the gap amplitude generally found in the literature and \emph{ii)}
the ZBCP is present and similar to the common ZBCP measured on HTCS\cite{alff98}.
In the metallic limit, the conductance is enhanced at low bias and
has a sharp point at zero bias indicative of a \( d \)-wave superconductor.

\section{Conclusion}

We realized and characterized planar junctions on the side of YBCO
single crystals. The direction of the injected current is controlled
within a few degrees in CuO\( _{2} \) planes. We showed that the interface
barrier can be changed from insulating to metallic by using different
preparation techniques. If the crystal is simply polished the resulting
interface will be insulating and single electron tunneling will be
the main mechanism of transport through the interface. On the other
hand, if the surface undergoes an ion polishing stage it will show
a metallic behaviour and Andreev reflections below the superconducting
transition. The mechanical stability of the junction  allows us  to perform
 Andreev
spectroscopy as a function of temperature. These investigations are
currently being undertaken and will be published elsewhere. They could provide
some experimental evidence of the ideas suggested by Deutscher~\cite{deutscher99}
by measuring different energy scales for pairing (tunneling spectroscopy)
and phase coherence (Andreev spectroscopy). Our results show that
under carefully established conditions the control of the surface
potential of HTCS can be achieved. Further research to understand HTCS
require the use of controlled interfaces. The technique presented here has 
yielded goog results. 
The authors would like to thank T. Fournier and P. Brosse-Marron for
technical support and O. Fruchart for initial help with the AFM work,
H. G. thanks B. Lussier for early supervision. 

\bibliographystyle{elsart-num}

\begin{thebibliography}{10}
\expandafter\ifx\csname url\endcsname\relax
  \def\url#1{\texttt{#1}}\fi
\expandafter\ifx\csname urlprefix\endcsname\relax\def\urlprefix{URL }\fi

\bibitem{bednorz86}
{Bednorz J.G.}, {M\"uller K.A.}, Z. Phys. B 64 (1986) 189.

\bibitem{orenstein00}
{Orenstein J.}, {Millis A.J.}, Advances in the {P}hysics of
  {H}igh-{T}emperature {S}uperconductivity, Science 288 (2000) 468.

\bibitem{loeser96}
{Loeser A.G.}, {Shen Z.-X.}, {Dessau D.S.}, {Marshall D.S.}, {Park C.H.},
  {Fournier P.}, {Kapitulnik A.}, Excitation {G}ap in the {N}ormal {S}tate of
  {U}nderdoped {BSCCO}, Science 273 (1996) 325.

\bibitem{pavuna00}
{Pavuna D.}, {Vobornik I.}, {Margaritondo G.}, Photoemission {E}xperiments on
  {H}igh-{T}c {S}uperconductors: {R}ecent {P}rogress and {S}ome {O}pen
  {Q}uestions, J. Supercond. 13~(5) (2000) 749.

\bibitem{renner98}
{Renner Ch.}, {Revaz B.}, {Kadowaki K.}, {Maggio-Aprile I.}, {Fischer O.},
  Observation of the {L}ow {T}emperature {P}seudogap in the {V}ortex {C}ore of
  {B}i2212, Phys. Rev. Lett. 80~(16) (1998) 3606--3609.

\bibitem{sharoni01}
{Sharoni A.}, {Koren G.}, {Millo O.}, Correlation of tunneling spectra with
  surface nano-morphology and doping in thin {YBCO} films, pr\'epublication
  Europhysics Lett. cond-mat/0103581.

\bibitem{convington97}
{Covington M.}, {Aprili M.}, {Paraoanu A.}, {Greene L. H.}, Observation of
  {S}urface-{I}nduced {B}roken {T}ime-{R}eversal {S}ymmetry in {YBCO} {T}unnel
  {J}unctions, Phys. Rev. Lett. 79~(2) (1997) 277--280.

\bibitem{fogelstrom97}
{Folgelstr\"om M.}, {Rainer D.}, {Sauls J. A.}, Tunneling into
  {C}urrent-{C}arrying {S}urface {S}tates of {H}igh-{T}c {S}uperconductors,
  Phys. Rev. Lett. 79~(2) (1997) 281--284.

\bibitem{Hu1994}
{Hu Chia-Ren}, Phys. Rev. Lett. 72~(10) (1994) 1526--1529.

\bibitem{alff98}
{Alff L.}, {Kleefisch S.}, {Schoop U.}, Andreev bound state in high temperature
  superconductors, Eur. Phys. J. B 5 (1998) 423--438.

\bibitem{lesueur91}
{Lesueur J.}, {Greene L.H.}, {Feldmann W.L.}, {Inam A.}, Zero bias anomalies in
  {YBCO} tunnel junctions, Physica C 191 (1992) 325--332.

\bibitem{aprili99}
{Aprili M.}, {Badica E.}, {Greene L. H.}, Doppler {S}hift of the {A}ndreev
  {B}ound {S}tates at the {YBCO} {S}urface, Phys. Rev. Lett. 83~(22) (1999)
  4630--4633.

\bibitem{krupke99}
{Krupke R.}, {Deutscher G.}, Anisotropic {M}agnetic {F}ield {D}ependence of the
  {Z}ero-{B}ias {A}nomaly on {I}n-{P}lane {O}riented [100] {YBCO}/{I}n {T}unnel
  {J}unctions, Phys. Rev. Lett. 83~(22) (1999) 4634.

\bibitem{walker99}
{Walker M.B.}, {Pairor P.}, Universal {W}idth for the {ZBA}, Phys. Rev. B
  60~(14) (1999) 10395--10399.

\bibitem{samokhin01}
{Samokhin K.V.}, {Walker M.B.}, Localized surface states in {HTSC} :
  alternative mechanism of zero-bias conductance peaks, Phys. Rev. B 64~(17)
  (2001) 172506.

\bibitem{deutscher99}
{Deutscher G.}, Coherence and single-particle excitations in the
  high-temperature superconductors, Nature 397 (1999) 410--412.

\bibitem{emery95}
{Emery V. J.}, {Kivelson S. A.}, Importance of phase fluctuations in
  superconductors with small superfluid density, nature 374 (1995) 434.

\bibitem{wolfbook}
{Wolf E.L.}, Principles of {E}lectron {T}unneling {S}pectroscopy, Oxford
  University Press, 1985.

\bibitem{hebard89}
{Hebard A.F.}, {Fleming R.M.}, {Short K.T.}, {White A.E.}, {Rice C.E.}, {Levi
  A.F.J.}, {Eick R.H.}, Ion beam thinning and polishing of {YBCO} films, App.
  Phys. Lett. 55~(18) (1989) 1915--1917.

\bibitem{chen97}
{Chen L.}, {Yang T.}, {Nie J.C.}, {Wu P.J}, {Huang M.Q.}, {Liu G.R.}, {Zhao
  Z.X.}, Enhancement of superconductivity by low energy {A}r ion milling in
  epitaxial {YBCO} thin films, Physica C 282--287 (1997) 657--658.

\bibitem{xavier94}
{Xavier P.}, {Fournier T.}, {Chaussy J.}, {Richard J.}, {Charalambous M.},
  Nonperturbative ion etching of {YBCO} thin films, J. Appl. Phys. 75~(2)
  (1994) 1219--1221.

\bibitem{carter01}
{Carter G.}, The {P}hysics and applications of ion beam erosion, J. Phys. D:
  Appl. Phys. 34 (2001) 1.

\bibitem{rzchowski2000}
{Rzchowski M.S.}, {Wu X.W.}, Bias dependence of magnetic tunnel junctions,
  Phys. Rev. B 61~(9) (2000) 5884--5887.

\bibitem{vandeveer97}
{van de Veerdonk R.J.M}, {Nowak J.}, {Meservey R.}, Current distribution
  effects in magnetoresistive tunnel junctions, Appl. Phys. Lett. 71~(10)
  (1997) 2839--2841.

\bibitem{pedersen67}
{Pedersen R.J.}, {Vernon Jr. F.L.}, Effect of film resistance on low-impedance
  tunneling measurements, Appl. Phys. Lett. 10~(1) (1967) 29--31.

\bibitem{deutscher91}
{Deutscher G.}, Proxymity effects with the cuprates, Physica C 185--189 (1991)
  216--220.

\bibitem{kupriyanov91}
{Kupriyanov M.Yu.}, {Likharev K.K.}, Towards the quantitative theory of the
  {H}igh-{T}c josephson junctions, IEEE trans. Magnetics 27~(2) (91)
  2460--2463.

\bibitem{tsai90}
{Tsai J.S.}, {Takeuchi I.}, {Tsuge H.}, {a}nomalous interface resistance
  between oxyde superconductors and noble metals, Physica B 165-166 (1990)
  1627--1628.

\bibitem{jing89}
{Jing T.W.}, {Wang Z.Z.}, {Ong N.P.}, Gold contacts on superconducting crystals
  of {YBCO} with very low contact resistivity, Appl. Phys. Lett. 55~(18) (1989)
  1912--1914.

\bibitem{mannhart99}
{Mannhart J.}, {Hilgenkamp H.}, {i}nterfaces involving complex superconductors,
  Physica C 317--318 (1999) 383--391.

\bibitem{koyanagi92}
{Koyanagi M.}, {Kashiwaya S.}, {Akoh H.}, {Kohjiro S.}, {Matsuda M.}, {Hirayama
  F.}, {Kajimura K.}, Study of {YBCO}/{A}u surface using {LTSTM}/{STS}, Jpn. J.
  Appl. Phys. 31~(11) (1992) 3525--3528.

\bibitem{cren00a}
{Cren T.}, {Roditchev D.}, {Sacks W.}, {Klein J.}, {Moussy J.-B.},
  {Deville-Cavellin C.}, {Lagu\"es M.}, Influence of {D}isorder on the {L}ocal
  {D}ensity of {S}tates in {H}igh-{T}c {S}uperconducting {T}hin {F}ilms, Phys.
  Rev. Lett. 84~(1) (2000) 147--150.

\end{thebibliography}

\newpage

\end{document}